\documentclass[12pt]{article}
\usepackage{amssymb}
\usepackage{amsmath}
\def\dd{{\rm d}}
\def\beq{\begin{equation}}\def\eeq{\end{equation}}
\def\bea{\begin{eqnarray}}\def\eea{\end{eqnarray}}
\def\half{{\textstyle\frac12}}\def\fourth{{\textstyle\frac14}}

\begin{document}

\title{Gravity and Matter in Causal Set Theory}

\author{Roman Sverdlov$^1$ and Luca Bombelli$^2$\\
\\$^1$ Physics Department, University of Michigan,
\\450 Church Street, Ann Arbor, MI 48109-1040, USA
\\$^2$ Department of Physics and Astronomy, University of Mississippi,
\\108 Lewis Hall, University, MS 38677-1848, USA}
\date{July 13, 2008}
\maketitle

\begin{abstract}
\noindent The goal of this paper is to propose an approach to the formulation of dynamics for causal sets and coupled matter fields. We start from the continuum version of the action for a Klein-Gordon field coupled to gravity, and rewrite it first using quantities that have a direct correspondent in the case of a causal set, namely volumes, causal relations, and timelike lengths, as variables to describe the geometry. In this step, the local Lagrangian density ${\cal L}(f;x)$ for a set of fields $f$ is recast into a quasilocal expression ${\cal L}_0(f;p,q)$ that depends on pairs of causally related points $p \prec q$ and is a function of the values of $f$ in the Alexandrov set defined by those points, and whose limit as $p$ and $q$ approach a common point $x$ is ${\cal L}(f;x)$. We then describe how to discretize ${\cal L}_0(f;p,q)$, and use it to define a causal-set-based action.
\end{abstract}

\noindent{\bf 1. Introduction}

\noindent A causal set \cite{BLMS} (for recent reviews, see Refs \cite{RS,FD,JH}) is a locally finite partially ordered set, i.e., a set $S$ with a partial order relation $\prec$ in which for every two elements $p$ and $q$, the interval, or Alexandrov set, $\alpha(p,q):= \{r \mid p \prec r \prec q\}$ is finite. The causal set approach to quantum gravity uses causal sets as the only fundamental structure for the description of the geometry of spacetime, and is based on the observation that, if we interpret the partial order as the set of causal relations and define the volume of a subset of a causal set to be simply the number of elements in it, we obtain a discrete version of all variables needed to describe a Lorentzian geometry. In the context of the continuum, it is relatively simple to see that a Lorentzian metric tensor field is equivalent to a light-cone structure and a volume element at every point, and in a causally well-behaved Lorentzian manifold the equivalence carries over to the causal structure. More precisely, the topology, differentiable structure, and conformal metric of any past and future-distinguishing spacetime are uniquely determined by the partial order defined by the causal relations among its points \cite{HKM,Mal}; a similar result holds when a countable, dense set of randomly chosen points in the manifold is used \cite{BoMe}. It is conjectured that, if the causal relations are known among points randomly chosen with finite density $\rho$, then the conformal structure can be approximately determined, up to variations on small scales compared to the point density, but in addition the volumes of spacetime regions (and thus the whole metric, up to a global scale factor) can be determined by counting points in them, with the same type of approximation. These kinematical considerations tell us that, although causal sets are not geometrical objects, and most of them do not correspond to Lorentzian geometries in any natural way, those that are ``manifoldlike" in an appropriate sense are likely to define an effectively unique large-scale geometry.

Formulating a reasonable dynamics for causal sets has remained an elusive goal. Various proposals have been made, of which the most studied one is the classical sequential growth dynamics of Rideout and Sorkin \cite{RiSo}, in which causal sets (``causets" for short) are grown one element at a time, with each new element appearing to the future of some subset of the existing set, the specific probabilities for the choice of that subset being characteristic of the model used. Thus, each model gives a probability distribution on the set of $N$-element causets after $N$ elements have been generated. In this approach, if one imposes suitably formulated discrete versions of general covariance and causality, each possible model is characterized by a sequence of parameters, which allows the whole set of models to be systematically studied. As a result, some properties of causal sets that arise from these models are known, in particular for the special type called ``transitive percolation" (or ``random graph orders" in the mathematics literature) \cite{TransPerc}, but also for more general models \cite{AshMc} and, although any causal set can be generated by almost any one of the models, there are indications that the probability that a manifoldlike causal set emerges goes to zero as $N\to \infty$ \cite{Bright}.

In this paper, we will follow a different approach to the dynamics of causal sets, which is in principle more straightforward: We seek a way to write down the usual continuum action for gravity, and possibly matter fields, where the metric has been replaced by causal relations and volumes and the additional fields expressed in terms that have a well-defined causal set correspondence, and then discretize the resulting expression. In practice, this task is not as simple as it sounds. First of all, while using the volume element as a variable can be thought of as simply identifying the occurrence of $\sqrt{-g}$ in the action, it would not be useful to think of the locally defined conformal structure, $|g|^{-1/d}\,g_{\mu\nu}$ in $d$ spacetime dimensions, which is tensorial rather than combinatorial in nature, as the variable representing the causal structure, for the purpose of discretizing the theory. Instead, we will use as variable the actual partial order $p \prec q$, or $q \in J^+(p)$, even in the continuum. This, however, will introduce some non-locality in the theory; we will elaborate on this point below. The difficulty involved in the inclusion of matter fields in this framework depends on the type of field considered, but the case of a scalar field is simple enough that we shall treat it first, even before the case of gravity itself; results for other fields will be discussed in separate articles \cite{Roman}. The final issue related to our approach consists in specifying how to use the action we obtain in a dynamical principle. Although fully addressing this issue is beyond the scope of this paper, we will include some remarks on it.

\newpage
\bigskip
\noindent{\bf 2. Our Program}

\noindent To illustrate the issues that arise when writing a form of the action for a set of fields that is meaningful in the causal set context, consider the example of a Klein-Gordon field $\phi$ minimally coupled to a metric $g_{\mu\nu}$ (considered as fixed, for now), with Lagrangian density\footnote{Although, strictly speaking, a Lagrangian density should include the volume element $\sqrt{-g}$, in this paper we prefer to include the latter in the integration that gives the action $S = \int_{\cal M}\dd^4x\,\sqrt{-g}\,{\cal L}$.}
\beq
{\cal L}_{\rm KG}(g_{\mu\nu},\phi,x) = \half\,
(g^{\mu\nu}\, \partial_\mu\phi\, \partial_\nu\phi-m^2\phi^2)\;. \label{kg}
\eeq
If our goal was simply to discretize the continuum theory defined by Eq (\ref{kg}), for example as a computational tool, we could introduce a regular lattice and convert the derivatives into finite differences using standard methods. In a fundamentally discrete theory this would not be a reasonable approach, because a regular lattice introduces extra structure in the theory and breaks its covariance. The only way to overcome this difficulty with the discretization process is to use a ``random lattice" \cite{BHS}; we will in fact view a causal set as a random discretization of a Lorentzian manifold $(M,g_{\mu\nu})$ for some purposes. On a random lattice, however, covariant versions of finite-difference Lagrangian densities typically lead to divergent expressions. For example, we could try to replace the derivatives in Eq (\ref{kg}) with a sum of terms of the type $(\phi(q)-\phi(p))^2/(\tau(p,q))^2$, where $\tau(p,q)$ is the length of the longest future-pointing timelike curve from $p$ to $q$, over all timelike ``nearest neighbors" $q$ of $p$ (spacelike distances would not carry over well to the causal set context). But in a random lattice in a Lorentzian spacetime there are no ``nearest neighbors", in general. Even in Minkowski space of any dimensionality $d \ge 2$, for any $p \in {\cal M}$ and $\epsilon > 0$ the set of points $q$ for which $\tau(p,q) < \epsilon$ is a region of infinite volume,\footnote{It is the region between the future light cone of $p$, given by $t = r$ if $p$ is the origin, and the hyperboloid $t = \sqrt{x^2+\epsilon^2}$. Since the surface area of a $(d-2)$-dimensional sphere is $c_{d-2}\,r^{d-2}$ for some positive constant $c_{d-2}$, the volume of the region is
$V = \int_0^\infty (\sqrt{r^2+\epsilon^2}-r)\,c_{d-2}\,r^{d-2}\,\dd r
= \int_0^\infty \big(\half\,\epsilon^2\,c_{d-2}\,r^{d-3} + O(\epsilon^4)\big)
\,\dd r = \infty$.} so there are infinitely many random lattice points within an arbitrarily small Lorentzian distance, and none of them minimizes that distance. Thus, since the terms we would be adding are all non-negative, any non-trivial, Lorentz-invariant definition of this type would give divergent results, and the divergence cannot be removed by some renormalization procedure without violating Lorentz invariance.

The way to avoid the above infinity in ${\cal L}_{\rm KG}$ is to restrict ourselves to summing the terms $(\phi(q)-\phi(p))^2/(\tau(p,q))^2$ only over pairs of points within some finite region around $x$. In the context of causal sets, the simplest kind of finite region we can define using the partial order is an Alexandrov set. Thus, we will try to reconstruct $g^{\mu\nu}\partial_\mu\phi\,\partial_\nu \phi(x)$ by solely looking at the behavior of $\phi$ in the interior of an Alexandrov set centered at $x$. It may at first appear that this cannot work, because choosing one particular Alexandrov set is equivalent to choosing a preferred frame at $x$, the one in which the axis of the Alexandrov set coincides with the $t$ direction, and our results will therefore not be covariant. However, what we are trying to do is determine the value of a scalar by using values of quantities defined in a particular frame, which is certainly possible. In fact, as we show in Sec 2, if $\phi$ is a differentiable function, then it is possible to find a small enough Alexandrov set $\alpha(p,q)$ such that
\beq
g^{\mu\nu}\, \partial_\mu \phi\, \partial_\nu \phi(x)
\approx A\,\frac{(\phi(q)-\phi(p))^2}{(\tau(p,q))^2}
+ \frac{B}{(\tau(p,q))^6} \int_{\alpha(p,q)}\dd^4y
\ (\phi(y)-\phi(p))^2\;, \label{lag1}
\eeq
where $A$ and $B$ are appropriately chosen coefficients; a corresponding expression for gravity is derived in Sec 3. This result can be seen as analogous to finding an expression for a scalar invariant of the Faraday tensor $F_{\mu\nu}$ in electromagnetic theory using a combination of the electric and magnetic fields in some reference frame, with appropriately chosen coefficients to make the result covariant.

There is one important difference between the electromagnetic example and our situation, however, which is that the quantities appearing in (\ref{lag1}) do not have simple transformation laws under changes in $\alpha(p,q)$, except in the limit of infinitesimal sets, in which the $\Delta\phi$ change linearly under Lorentz transformations. As a consequence, the right-hand side is only covariant, and in fact only approaches the left-hand side, in the limit as $p$, $q \to x$, or $\alpha(p,q) \to \{x\}$. Notice that this is not the same as the $\tau(p,x)$, $\tau(x,q) \to 0$ limit. The points $p$ and $q$ can be arbitrarily close to $x$, or to each other, in the latter sense and still have coordinates that differ by arbitrarily large amounts in any given reference frame, just by being sufficiently close to each other's light cone. Therefore, even smooth functions may not have well-defined limits in the sense of the Lorentzian distance.

Once we have found an Alexandrov-set-based expression such as that in (\ref{lag1}) for a Lagrangian density that converges to the right value as $p$, $q \to x$, the remaining steps are: (i) Choose a specific pair $(p,q)$ for each point $x$; (ii) Write down a discretized version of the Lagrangian (\ref{lag1}) where $x$ is an element of a causal set that may or may not be manifoldlike; and (iii) Sum the contributions from all $x$ to the action. The tricky thing about the first step is how to achieve the desired result in a covariant way; in Sec 4 we will describe one way of doing this, which consists in choosing the interval in which the field varies more slowly, in an appropriate sense. Part of the second step can be taken care of in an obvious way: we consider first the case of a causal that is manifoldlike, and embedded uniformly in a Lorentzian manifold with density $\rho$, in which case we can replace generic manifold points by embedded causal set elements, and integrals over Alexandrov sets by sums over elements contained in the corresponding intervals, each element with a weight $\rho^{-1}$; since the resulting expression only contains quantities that are meaningful in a general causal set, we simply use the same one in the non-manifoldlike case. What is not so obvious in the causal set case is what to replace the $p$, $q \to x$ limit by. That limit cannot be taken in the same way as in the continuum for the simple reason that the set is locally finite, and we restrict ourselves to using intervals with at least a certain minimum number $N$ of elements, to minimize the effects of statistical fluctuations. The third step is then essentially trivial.

\bigskip
\noindent{\bf 3. Scalar Field}

\noindent Consider a massive Klein-Gordon scalar field in a curved $d$-dimensional spacetime, $d\ge2$, with Lagrangian density written in the usual form (\ref{kg}). We wish to derive an alternative expression for ${\cal L}_{\rm KG}(g_{\mu\nu},\phi,x)$ in which the dependence on $g_{\mu\nu}$ is expressed only through the causal structure and volume element. The action for gravity itself will be treated in the next section, but here we will lay some of the groundwork for that discussion as well.

We are interested in calculating integrals of expressions involving $\phi(x)$ over small Alexandrov sets $\alpha(p,q)$, which in fact will approach a point in the continuum. We can therefore assume that $\alpha(p,q)$ is a convex normal neighborhood of its midpoint and expand all tensor fields inside $\alpha(p,q)$ in Riemann normal coordinates $x^\mu$ based at that point, with the vector $\partial/\partial x^0|_0$ pointing along the geodesic from $p$ to $q$. For the scalar field we simply expand
\beq
\phi(x) - \phi(p) = \partial_\mu\phi|_0\,x^\mu + o(x)\;, \label{deltaphi}
\eeq
for small $x^\mu$, where ``little o" is defined as usual by the property that $o(x^k)$ vanishes faster than $k$-th powers of the coordinates as $x^\mu \to 0$. We can similarly expand all geometrical quantities. The metric itself, in Riemann normal coordinates, can be written in the form
\beq
g_{\mu\nu} = \eta_{\mu\nu} + {\textstyle\frac13}\,
R_{\mu\rho\nu\sigma}(0)\,x^\rho x^\sigma + o(x^2)\;,
\eeq
whose volume element is
\beq
\dd V(x) = |g(x)|^{1/2}\,\dd^dx = \big(1+O(x^2)\big)\,\dd^dx\;,
\eeq
and we can define an $h$ so that the coordinates of the endpoints $p$ and $q$ are given by $(\mp h,0,...,0)$ up to terms of order $h^2$ or higher; specifically, we choose
\beq
h:= \half\,\tau(p,q)\;.
\eeq

What we wish to do is calculate the integral on the right-hand side of (\ref{lag1}),
\beq
\int_{\alpha} (\phi(x)-\phi(p))^2\, \dd V_x
= (\partial_\mu \phi\,\partial_\nu \phi)\big|_0 \int_{\alpha_0}
(x^\mu - p^\mu)(x^\nu - p^\nu )\,\dd^dx + o(h^{d+2})\;, \label{int6}
\eeq
where $\alpha_0$ is the Alexandrov set of the points $(\mp\,\tau/2,0,...,0)$ with respect to the flat metric $\eta_{\mu\nu}$, in terms of derivatives of $\phi$. Since $\alpha_0$ is just a pair of cones of height $h$ in Minkowski space over a $(d-1)$-dimensional ball of radius $h$ and volume
\beq
V({\rm B}^{d-1},h) = c_{d-1}\,h^{d-1},\quad {\rm with}\quad
c_d = \frac{2\,\pi^{d/2}}{d\,\Gamma(d/2)}
\eeq
(for example, $c_1 = 2$, $c_2 = \pi$, $c_3 = \frac43\,\pi$, $c_4 = \half\,\pi^2$, $c_5 = \frac8{15}\,\pi^2$, ...), the volume of $\alpha_0$ is simply
\beq
V_0:= \int_{\alpha_0} \dd^d x = 2\cdot\frac{c_{d-1}\,h^d}{d} = k_d\,\tau^d\;,
\qquad k_d:= \frac{c_{d-1}}{2^{d-1}\,d}\;. \label{V0}
\eeq

Expanding the product in the integrand on the right-hand side of (\ref{int6}), and noticing that integrals of odd functions over symmetric domains such as $\alpha_0$ vanish, we can rewrite (\ref{int6}) as 
\bea
& &\int_{\alpha} (\phi(x)-\phi(p))^2\, \dd V_x
=\ \sum_{\mu=0}^{d-1} (\partial_\mu\phi|_0)^2
\bigg(\int_{\alpha_0} (x^\mu)^2\, \dd^dx
+ (p^\mu)^2 \int_{\alpha_0} \dd^d x \Big)\nonumber\\
& &\kern112pt=\ \big(I_{d,0}+\fourth\,k_d\big)\, \tau^{d+2}\,(\partial_0\phi)^2
+ I_{d,1}\,\tau^{d+2} \,(\partial_k\phi\,\partial_k\phi)\;, \label{int1}
\eea
where a summation over $k = 1, ..., d-1$ is implied, and we have defined the integrals
\bea
& &I_{d,0}:= \frac1{\tau^{d+2}} \int_{\alpha_0} (x^0)^2\,\dd^dr
= \frac{c_{d-1}}{2^d\, d\, (d+1)\,(d+2)}\;, \label{Id0}\\
& &I_{d,1}:= \frac1{\tau^{d+2}} \int_{\alpha_0} (x^1)^2\,\dd^dr
= \frac{c_{d-2}\,J_{d+1}}{2^{d+1}\,d\,(d+2)}\;, \label{Id1}
\eea
with
\beq
J_d:= \int_{-\pi/2}^{\pi/2} (\cos\theta)^d\,\dd\theta \label{Id}
\eeq
(thus, $J_1 = 2$, $J_2 = \half\,\pi$, $J_3 = \frac43$, $J_4 = \frac38\,\pi$, $J_5 = \frac{16}{15}$, ...). To calculate the value of $I_{d,0}$, we can use the fact that the Alexandrov set $\alpha_0$ is symmetric under $t \mapsto -t$, and each of the two cones it consists of can be sliced into $t$ = constant hypersurfaces that are simply $(d-1)$-balls of radius $h-|t|$. To obtain the value of $I_{d,1}$, we further slice each ball into $x_1$ = constant discs of radius $\sqrt{(h-|x_0|)^2-(x_1)^2}$.

We can now use Eqs (\ref{lag1}), (\ref{deltaphi}) and (\ref{int1}) to express the derivatives that enter ${\cal L}_{\rm KG}$ in terms of Alexandrov-set-based quantities. For the time derivative, (\ref{deltaphi}) gives us
\beq
\partial_0\phi = \frac{\phi(q)-\phi(p)}{\tau}
\eeq
(here and in the following we are neglecting higher-order terms in $\tau^{-1}$), and for the spatial derivatives, substituting this last result into (\ref{int1}), we get that
\bea
\partial_k\phi\,\partial_k\phi
= \frac{1}{I_{d,1}\,\tau^{d+2}} \int_{\alpha(p,q)} (\phi(x)-\phi(p))^2\,\dd^dx
- \frac{I_{d,0}+\fourth\,k_d}{I_{d,1}}\, \frac{(\phi(q)-\phi(p))^2}{\tau^2}\;.
\label{phis}
\eea
These last two equations allow us to write the kinetic term in ${\cal L}_{\rm KG}$ as
\bea
& &g^{\mu\nu}\,\partial_\mu\phi\,\partial_\nu\phi 
= \partial_0\phi\,\partial_0\phi - \partial_k\phi\,\partial_k\phi
\label{phil} \nonumber\\
& &= \bigg(1+\frac{I_{d,0}+\fourth\,k_d}{I_{d,1}} \bigg)\,
\frac{(\phi(q)-\phi(p))^2}{\tau^2} - \frac{1}{I_{d,1}\,\tau^{d+2}}
\int_{\alpha(p,q)} (\phi(x)-\phi(p))^2\, \dd^dx\;. \label{gphiphi}
\eea
On the other hand, if we choose to use the volumes of the Alexandrov set instead of $\tau$ to represent the separation between $p$ and $q$, then (\ref{V0}) gives us
\beq
\tau = \bigg(\frac{V}{k_d}\bigg)^{\!1/d}\;,
\eeq
which, upon substitution into (\ref{phil}) gives
\bea
& &g^{\mu\nu}\,\partial_\mu\phi\,\partial_\nu\phi
= \bigg(1+\frac{I_{d,0}+\fourth\,k_d}{I_{d,1}} \bigg)\,
\bigg(\frac{V}{k_d}\bigg)^{\!2/d}\,(\phi(q)-\phi(p))^2
\nonumber\\ & &\kern80pt
-\ \frac{1}{I_{d,1}}\,\bigg(\frac{V}{k_d}\bigg)^{\!1+2/d}
\int_{\alpha(p,q)} (\phi(x)-\phi(p))^2\, \dd^dx\;.
\eea

Thus, in four dimensions (for concreteness, in which case $k_4 = \pi/24$, $I_{4,0} = \pi/(2^5\cdot3^2\cdot5)$, $I_{4,1} = \pi/(2^4\cdot3^2\cdot5)$), the Alexandrov set version of the scalar field Lagrangian density (\ref{kg}) is
\bea
& &{\cal L}_{\rm KG}(\prec,|g|,\phi;p,q) \label{lag2}\\
& &= 3\,\frac{\big(\phi(q)-\phi(p)\big)^2}{\tau^2(p,q)}
- {\frac{360}{\pi\,\tau^{6}(p,q)}}
\int_{\alpha(p,q)}\big(\phi(x)-\phi(p)\big)^2\,\dd^4x
+ \half\,m^2\,\phi^2(p)\;,\qquad
\nonumber
\eea
up to $o(1)$ terms, where, in the continuum context, we view the proper time as a function of the causal structure and volume element, obtained from Eq (\ref{V0}),
\beq
\tau(p,q) = 2\,\bigg(\frac{V(p,q)}{k_d} \bigg)^{1/d}\;.
\label{tau}
\eeq

Finally, the scalar field action in terms of $\phi$, Alexandrov sets, and the volume element is obtained by taking the local limit and integrating over manifold points,
\bea
& &{\cal L}_{\rm KG}(\prec,|g|,\phi;x)
= \lim_{p,\,q\,\to\, x\atop p\prec x \prec q}\,
{\cal L}_{\rm KG}(\prec,|g|,\phi;p,q) \label{limit}\\
& &S_{\rm KG}
= \int_{\cal M} \dd^dx\,|g|^{1/2}\,
{\cal L}_{\rm KG}(\prec,|g|,\phi;x)\;.
\eea
Notice that the quasilocal expression (\ref{lag2}), viewed as a function of $(\prec,|g|,\tau)$, without using (\ref{tau}) to eliminate $\tau$, would in a sense resemble more the original ${\cal L}_{\rm KG}(g_{\mu\nu},\phi,x)$, with derivatives expressed by terms containing lengths rather than volumes. However, since the continuum dynamics uses the local Lagrangian density, and in the limit (\ref{limit}) the relationship (\ref{tau}) always holds, the set of variables $(\prec,|g|,\tau)$ is a redundant one, and this point of view makes no practical difference, even for the quantum theory. But in the discretized theory that is no longer true. Even for manifoldlike causal sets the relationship between $\tau(p,q)$ and the length $\ell(p,q)$ of the longest chain between $p$ and $q$ is not straightforward. Even for small distances, when those two quantities are approximately proportional in the mean,
\beq
\tau(p,q) = \kappa_d\,\langle\ell(p,q)\rangle\;, \label{timelike}
\eeq
up to curvature corrections, the coefficient $\kappa_d$ is dimension-dependent and not known analytically \cite{BrGr}. Most importantly, the action needs to be extended to non-manifoldlike causal sets, for which $\ell(p,q)$ as a variable is independent of the cardinality of the interval between $p$ and $q$.

\bigskip
\noindent{\bf 4. Gravity} 

\noindent We now wish to write down a similar form for the Einstein-Hilbert action for gravity,
\beq
S_{\rm EH}(g_{\mu\nu}) = \int_{\cal M}R\,\dd V_x\;,
\eeq
again in $d \ge 2$ spacetime dimensions. From Refs \cite{Myr,GibSol} we know that the generalization of the expression (\ref{V0}) for the volume of an Alexandrov set $\alpha$ of height $\tau$ to curved spacetime is
\beq
V(\tau) = k_d\,\tau^d \big[1+(a_d\,R+b_d\,R_{00})\,\tau^2 + o(\tau^2)\big]\;,
\label{V}
\eeq
where $k_d$ was defined in (\ref{V0}), the curvature is evaluated at the midpoint of $\alpha$,
\beq
a_d = \frac{d}{24\,(d+1)\,(d+2)}\;, \qquad b_d = \frac{d}{24\,(d+1)}\;,
\eeq
and the $o(\tau^2)$ terms contain higher powers of the curvature tensor.
This equation can be rewritten as
\beq
\frac{V(\tau)}{k_d\,\tau^d} - 1 = (a_d\,R + b_d\,R_{00})\,
\tau^2 + o(\tau^2)\,, \label{rewrite}
\eeq
where the left-hand side contains the kinds of variables we want to use in the causal set context, while the right-hand side contains $R$, the Lagrangian density, and $R_{00}$. Therefore, the idea is to use (\ref{rewrite}) to obtain two equations involving the same curvature components, and solve them for $R$. This will involve applying (\ref{rewrite}) to different Alexandrov sets expressed in the same coordinates, so we start by writing it in an equivalent covariant form, in terms
of the endpoints $p$ and $q$ of $\alpha$,
\beq
\frac{V(p,q)}{k_d\,\tau^d(p,q)} - 1 = (a_d\,R\,g_{\mu\nu} + b_d\,R_{\mu\nu})\,
(q^\mu - p^\mu)\, (q^\nu - p^\nu) + o(\tau^2)\,. \label{one}
\eeq

One of our two equations is (\ref{rewrite}) itself, and we obtain the other one by writing it for an Alexandrov set $\alpha(p,x)$ and integrating over all $x \in \alpha(p,q)$. To leading order, this gives
\bea
& &\int_{\alpha(p,q)} \bigg(\frac{V(p,x)}{k_d\,\tau^d(p,x)} - 1\bigg)\,\dd^dx
= (a_d\,R\,\eta_{\mu\nu} + b_d\,R_{\mu\nu})
\int_{\alpha(p,q)} (x^\mu - p^\mu)\, (x^\nu - p^\nu)\, \dd^dx \label{two}\\
& &=\ a_d\,R \int_{\alpha(p,q)}
\big[(x^0)^2 - x^k\,x^k + p_\mu\, p^\mu\big]\,\dd^dx
+ b_d \int_{\alpha(p,q)} \big[R_{00}\,(x^0-p^0)^2 + R_{ii}\,(x^i)^2\big]\,\dd^dx
\nonumber\\
& &=\ \Big\{a_d\,R\, \Big[I_{d,0} - (d-1)\,I_{d,1}+\fourth\,k_d\Big]
+ b_d\, \Big[R_{00}\,(I_{d,0}+\fourth\,k_d) + \sum\nolimits_iR_{ii}\,I_{d,1}\Big]
\Big\}\,\tau^{d+2}\;. \nonumber\\
& &=\ \Big\{ R\, \Big[ a_d \Big(I_{d,0} - (d-1)\,I_{d,1} + \fourth\,k_d \Big)
- b_d\,I_{d,1} \Big] + R_{00}\,b_d\, \Big(I_{d,0} + I_{d,1} + \fourth\,k_d \Big)
\Big\}\,\tau^{d+2}\;,
\nonumber\eea
where in the last step we have used $\sum_iR_{ii} = -R + R_{00}$. The two Eqs\ (\ref{rewrite}) and (\ref{two}) form a linear system for the unknowns $R$ and $R_{00}$. The determinant of the coefficients of those variables is
\beq
D = (a_d\, d + b_d\, I_{d,1})\,\tau^{d+2}\;,
\eeq
and when we solve the system for $R = {\cal L}_{\rm EH}$ we obtain the quasilocal expression
\beq
{\cal L}_{\rm EH}
= \frac1D\, \bigg\{ \bigg(\frac{V(\tau)}{k_d\,\tau^d} - 1\bigg)\,
\big(I_{d,0} + I_{d,1} + \fourth\,k_d\big)\,\tau^d
- \int_{\alpha(p,q)} \bigg(\frac{V(p,x)}{k_d\,\tau^d(p,x)} - 1\bigg)\,\dd^dx
\bigg\}\,. \label{grav}
\eeq

An important difference between this expression and the corresponding one for the scalar field, ${\cal L}_{\rm KG}$, is that now $V(p,q)$ and $\tau(p,q)$ are not equivalent variables, and therefore we cannot consider $(\prec,|g|,\tau)$ a redundant set as we did in the scalar field Lagrangian. Therefore, strictly speaking, the Lagrangian ${\cal L}_{\rm EH}$ in (\ref{grav}) is not expressed in terms of volumes and causal relations alone, since we need timelike lengths to evaluate it. While it is possible that an expression purely in terms of $(\prec,|g|)$ can be found, in this work we will use the above ${\cal L}_{\rm EH}$, since in the causal set context there is a way to estimate timelike lengths independently of Alexandrov set volumes.
\newpage
\bigskip
\noindent{\bf 5. Discretization of the Lagrangian}

\noindent In the previous two sections we showed how to rewrite the action for a Klein-Gordon scalar field and for general relativity using just the causal structure $\prec$, the volume element $|g|$, and the Lorentzian distance $\tau$ as geometric variables. The general form of the action was that of an integral over $\cal M$ of a Lagrangian density for a set of fields $f$ (which include the matter fields as well as the geometrical variables) obtained as a limit ${\cal L}(f;x) = \lim_{p,\,q\to x} {\cal L}_0(f;p,q)$ of an expression that depends on $f$ inside an Alexandrov set $\alpha(p,q)$ centered at $x$. The quasilocal Lagrangian density ${\cal L}_0(f;p,q)$ can be readily discretized, as we will show in this section; in the next section we will propose an approach to defining a discrete version of the limiting procedure needed to build the action from the quasilocal Lagrangian density.

If we assume that the relationship between a continuum geometry $({\cal M}, g_{\mu\nu})$ and a causal set $\cal C$, when the latter is ``manifoldlike", is that $\cal C$ appears as a uniformly random set of points embedded in $({\cal M}, g_{\mu\nu})$ with density $\rho$, then the discrete version ${\cal L}(f;p,q)$ can be obtained simply by taking $p$ and $q$ to be embedded causal set elements with $p \prec q$, replacing the Alexandrov set $\alpha(p,q)$ in $\cal M$ by the interval between $p$ and $q$ in $\cal C$, and converting every integral over an Alexandrov set into a sum over interval elements, each carrying a weight $\rho^{-1}$, with any matter fields being evaluated again using the embedded points. The specific way in which this last step is carried out depends on the type of fields under consideration, but in the case of a scalar field it is obvious, and for a Klein-Gordon field Eq (\ref{gphiphi}) gives
\bea
& &{\cal L}_{\rm KG}(\prec,\tau,\phi;p,q) \label{KGd}\\
& &= \bigg(1+\frac{I_{d,0}+\fourth\,k_d}{I_{d,1}} \bigg)\,
\frac{(\phi(q)-\phi(p))^2}{2\,\tau^2(p,q)} - \frac{\rho^{-1}}{2\,I_{d,1}\,
\tau^{d+2}(p,q)}\sum_{p\prec x\prec q}(\phi(x)-\phi(p))^2
- \half\,m^2\,\phi(p)^2\;. \nonumber
\eea
where $d$ and $\rho$ are considered parameters, and we obtain different versions depending on whether we express $\tau(p,q)$ in terms of the cardinality $V(p,q)$ of the interval $\alpha(p,q)$ using (\ref{tau}) or in terms of the length $\ell(p,q)$ of the longest chain between $p$ and $q$ using Eq (\ref{timelike}).

We can similarly discretize the gravitational field Alexandrov-set-dependent Lagrangian density, and for the Einstein-Hilbert action Eq (\ref{grav}) gives
\bea
& &{\cal L}_{\rm EH}(\prec,\tau;p,q) \nonumber\\
& &= \frac1D\, \bigg\{ \bigg(\frac{V(\tau)}{k_d\,\tau^d} - 1\bigg)\,
\big(I_{d,0} + I_{d,1} + \fourth\,k_d\big)\,\tau^d
- \rho^{-1}\sum_{p\prec x\prec q} \bigg(\frac{V(p,x)}{k_d\,\tau^d(p,x)} - 1\bigg)
\bigg\}\,, \label{EHd}
\eea
where $d$ and $\rho$ are still considered parameters, but in this case we need to think of $\tau(p,q)$ as expressed in terms of $\ell(p,q)$ using $\tau = \kappa_d\,\ell$, the relationship that holds in a manifoldlike causal set.

From general facts about causal sets, we know that the manifoldlike ones ---the ones ``on the manifold shell", so to speak--- are a tiny fraction of the total, but they are the only ones we know how to use for the purpose of checking the correctness of the dynamics, so we will just extend the expressions just obtained for the manifoldlike case to all causal sets; this is similar to what we do, successfully, in the path integral formulation of particle mechanics and flat spacetime field theory, where most of the histories that contribute are nowhere differentiable and nothing like the ones we consider classically.

\newpage
\bigskip
\noindent{\bf 6. The discrete action}

\noindent We will now address the issue of selecting the ``right" Alexandrov set around each point, for the purpose of summing over points and computing our actions. There are two difficulties here: (1) If we select an Alexandrov set with too few points, then random fluctuations will no longer make our approximation reliable, and (2) If the axis of Alexandrov set is too close to the light cone, then despite the fact that the Lorentzian distance between the two endpoints of Alexandrov set might be small, the difference between the coordinate values of these points may still be arbitrarily large, hence arbitrarily large fluctuations of the fields are possible. The way to address the first issue is simple: we have to set the condition that the number of elements in the Alexandrov set be greater than some fixed number, or equivalently, $V(\alpha(p,q)) > V_1$.

As far as the second condition is concerned, this one is a little bit more tricky, because the notion of being ``sufficiently far away from the boundaries of the light cone" is not relativistically covariant. But then the question is this: why is the fact that the notion is not covariant not a problem in practice when talking about slowly-varying fields? To try to answer this, we can imagine a situation where things don't work near the light cone of a point $p$. For example, a planet that is a million kilometers away from $p$ moves parallel to the $t$ axis, and eventually crosses the light cone of $p$. Consider a point $q$, that lies on the path of that planet and is {\em not\/} on the light cone of $p$ but to its future, so close to it that its Lorentzian distance to $p$ is very small. We can now make a Lorentz boost so that $p$ and $q$ both lie on the new $t$ axis. In this case, the picture we will see is that $q$ is close to $p$ coordinatewise, but this didn't help our situation because the planet was moving so fast in the new coordinates that despite the fact that very little time had elapsed between $p$ and $q$, it still ``had time" to hit $q$ despite having been very far away from $p$. 

Thus, what we see is that the question we need to address is really {\em not\/} one about relativity and covariance, but rather how come the scenario just described is not a problem. In fact, the same issue arises in Newtonian physics and Galilean invariance, the only difference being that in that case the allowed range of velocities is infinite. In both cases, it would seem that we should run into the problem that, since the invariance group is not compact, if a planet were to pick a random velocity, it would pick with probability 1 one outside any given range of velocities. In special relativity, this means that if things were truly random, most planets would have been moving arbitrarily close to the speed of light, thus causing the disturbances we just described. So the fact that we see our spacetime as relatively continuous, without being disturbed by intruding planets, is an expression of the simple fact that in our universe, because of interactions, motion is not totally random, and the behavior of the matter in it selects preferred frames. What we have to do is find a way of selecting a preferred frame, in our case a preferred Alexandrov set, based on the behavior of fields near each point, and if we are able to do this covariantly we will not violate relativity.

One way to select an Alexandrov set based on the behavior of the fields is to choose the one that has the smallest fluctuation in the Lagrangian density ${\cal L}$, with the constraint already mentioned that its volume should be greater than some fixed $V_1$. In order to do that, we have to be able to measure the fluctuations of ${\cal L}$. If ${\cal L}$ was a local function, then we could just find the point with the largest ${\cal L}$ and the one with the smallest ${\cal L}$ in a given Alexandrov set, and look at the difference between the two values. But our Lagrangian densities ${\cal L}$ are quasilocal functions of whole Alexandrov sets. Thus, instead of using points, we have to use smaller Alexandrov sets embedded into our large Alexandrov set. But then again we have to be careful: if the ``small" Alexandrov sets are too small, then due to the fluctuation effects the results would no longer be reliable. So we have to again impose a lower bound on the volume of the ``smaller" Alexandrov sets we are considering, which we will call $V_2$. This means that the lower bound for the ``large" Alexandrov set we are looking at would have to be ``larger" than we had expected before: in addition to $V_1$ being large enough for the ``large" Alexandrov set to be reliable, it should also be large enough for there to exist $V_2 \ll V_1$ such that the Alexandrov sets of volume $V_2$ will also be large enough to be reliable.

Let's make what we have just said more precise. Suppose we have a set of fields $f$, which includes both the gravitational field, $\prec$, as well as any relevant matter fields. Our expression for the Lagrangian is a function both of $f$ and an Alexandrov set $\alpha(p,q)$: 
\beq
{\cal L} = {\cal L}(f;\alpha(p,q))\;.
\eeq
The fluctuation of ${\cal L}$ over the Alexandrov set based at points $p$ and $q$ is then given by 
\bea
& &{\rm fluct}({\cal L},f;V_2,\alpha(p,q))
= \max \{ {\cal L}(f;\alpha(r,s)) \mid
r,\,s \in \alpha(p,q);\ V(\alpha(r,s)) \geq V_2 \} \nonumber\\
& &\kern124pt-\ \min \{ {\cal L}(f;\alpha(r,s)) \mid r,\,s \in
\alpha(p,q);\ V(\alpha(r,s)) \geq V_2  \}\;.\quad
\eea
For any point $p$, define the set of points $q$ such that the Alexandrov set $\alpha(p,q)$ minimizes the above fluctuation, subject to the constraint that its volume be greater than $V_1$:
\bea
& &Q({\cal L},f,p, V_1, V_2) = \{ q\succ p \mid
\forall r\succ p\ \hbox{such that}\ (V(\alpha(p,r)) \geq V_1\,,\nonumber\\
& &\kern144pt{\rm fluct}({\cal L},f;V_2,\alpha(p,r))
\geq {\rm fluct}({\cal L},f;V_2, \alpha (p,q)) \}\;.\quad
\eea
Notice that in practice we expect $Q$ to consist of a single point, in most cases.

Finally, the Lagrangian density at a point $p$ is given by the average of the values determined by the set of all ``smoothest" Alexandrov sets; in other words, we are averaging over all Alexandrov sets $\alpha(p,q)$ with $q \in Q({\cal L},f,p, V_1, V_2)$,
\beq
{\cal L}(f,p,V_1,V_2) = \frac{\sum_{q \in Q({\cal L},f,p, V_1, V_2)}
{\cal L}(f;\alpha(p,q))}{\sharp\{Q({\cal L},f,p,V_1,V_2)\}}\;,
\eeq
where $\sharp\{...\}$ stands for the cardinality of a set.

\bigskip
\noindent{\bf 6. Conclusions}

\noindent In this paper we have obtained expressions for the action for a Klein-Gordon scalar field and for gravity in the causal set approach to quantum gravity. This was achieved in three steps. The first one, involving just the continuum theory for a set of fields $f$, consisted in finding a quasilocal Lagrangian density ${\cal L}_0(f;p,q)$ defined for pairs of causally related points $p \prec q$, such that its limit as $p$, $q \to x$ is the desired local density ${\cal L}(f;x)$, and the metric appears in ${\cal L}_0$ only through the causal structure $\prec$, the volume element $|g|$, and the Lorentzian distance $\tau$. The second step was the discretization; this is logically the pivotal point in the process, but in practice it turned out to be entirely straightforward, at least for the theories considered here. The key feature of our approach is that the discretization is carried out at the level of the quasilocal, Alexandrov set version ${\cal L}_0$ of the Lagrangian density, because it is only there that a quantity governing the dynamics depends on the spacetime geometry only through the triple $(\prec,|g|,\tau)$, each element of which has a causal set counterpart. The third step was the limiting procedure which gave us a discrete version of the local Lagrangian density, whose sum over causal set elements is the discrete action.

The expressions we found do not constitute yet a full proposal for the discrete dynamics of causal sets and matter fields, for various reasons, in addition to the fact that we have not spelled out how they are to be used. For one, they depend on two integer numbers $V_1$ and $V_2$, for which we have not specified how to choose values. In addition to those parameters, they also include some ambiguities, including (i) the possibility of replacing $\tau(p,q)$ (but not $\tau(x,y)$) in some or all of the places where it appears in (\ref{KGd}) and (\ref{EHd}) by $V(p,q)$ using Eq (\ref{tau}), and (ii) different choices for how to extend the action off the ``manifold shell". It is also possible that our expressions are simply an intermediate step in the derivation of other, more fundamental expressions in causal set terms. Finally, we should keep in mind that in the continuum one normally has to add boundary terms to the Einstein-Hilbert part of the action to satisfy certain criteria such as finiteness for asymptotically flat spacetimes and additivity, and those additional terms may change the structure of the causal set version of the action once they are taken into account.

On the other hand, the fact that the dynamics of matter fields can be treated together with gravity along the same lines is a definite advantage of our approach. The approach raises interesting questions, such as whether the discrete action, obtained assuming the causal set was manifoldlike, will lead to such causal sets being dynamically preferred. The role played by Eq (\ref{tau}) in our construction, as well as the fact that we expect the discrete Lagrangian density ${\cal L}$ to approach a value independent of $V_1$ and $V_2$ as their values decrease (aside from ``statistical fluctuations") also singles out one of the possible necessary conditions we can use to identify manifoldlike causal sets. If a causal set satisfies the property that, as two elements $p \prec q$ approach a common element $x$, the longest chain between them and the cardinality of the interval they define are related by Eq (\ref{tau}) for some $d$ that is independent of $x$ (once statistical fluctuations are taken into account), we may not be able to conclude that the causal set is manifoldlike, but the different versions of the actions (\ref{KGd}) and (\ref{EHd}) obtained from the continuum will coincide.

\end{document}